\pdfoutput=1
\documentclass[journal=jpclcd,manuscript=article,layout=twocolumn]{achemso}

\usepackage{chemformula} 
\usepackage[T1]{fontenc} 
\usepackage{mathptmx}
\usepackage{etoolbox}
\usepackage{amsmath}
\usepackage{amssymb}

\usepackage{braket}
\usepackage{color}
\usepackage{ulem}
\usepackage{accents}
\usepackage{chemscheme}
\usepackage{graphicx}
\usepackage{multirow}

\definecolor{darkgreen}{RGB}{40,110,5}

\usepackage{pgfplotstable}
\usepackage{rotating}
\usepackage{tabularx,booktabs}
\usepackage{adjustbox}
\usepackage{sidecap}
\usepackage{gensymb}
\usepackage{float}

\DeclareUnicodeCharacter{2212}{-}

\newcommand{\onlinecite}[1]{\hspace{-1 ex} \nocite{#1}\citenum{#1}}

\author{Leonel Varvelo}
 \altaffiliation{Co-first author: These authors contributed equally and authorship order was determined by a coin-toss. All authors agree that these authors may list themselves in either order for their CV/Resume and other purposes.}
 \affiliation{Department of Chemistry, Southern Methodist University, PO Box 750314, Dallas, TX, USA}
\author{Jacob K. Lynd}%
\altaffiliation{Co-first author: These authors contributed equally and authorship order was determined by a coin-toss. All authors agree that these authors may list themselves in either order for their CV/Resume and other purposes.}
\affiliation{Department of Chemistry, Southern Methodist University, PO Box 750314, Dallas, TX, USA}
\author{Brian Citty}%
\affiliation{Department of Chemistry, Southern Methodist University, PO Box 750314, Dallas, TX, USA}
\author{Oliver K\"{u}hn}
\affiliation{%
Institute of Physics, University of Rostock, Albert-Einstein-Str. 23-24, 18059 Rostock, Germany
}%
\author{Doran I. G. B. Raccah}
\affiliation{Department of Chemistry, Southern Methodist University, PO Box 750314, Dallas, TX, USA}
\email{doranb@smu.edu}
\title[An \textsf{achemso} demo]
  {Formally Exact Simulations of Mesoscale Exciton Diffusion in a Photosynthetic Aggregate}

\abbreviations{IR,NMR,UV}

\begin{document}
\sloppy 


\begin{tocentry}

\includegraphics{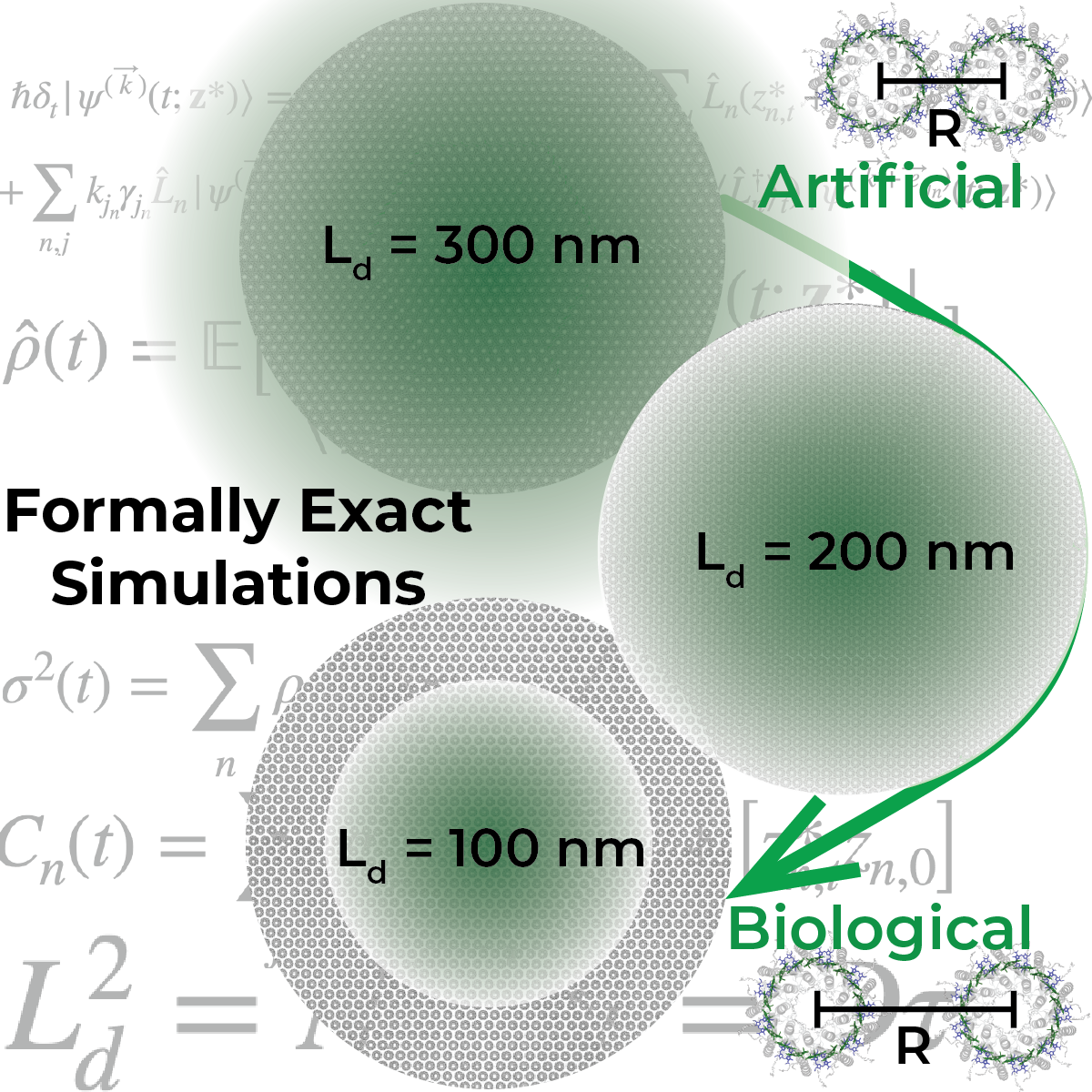}





\end{tocentry}


\begin{abstract}
\label{LH2:abs}
  The photosynthetic apparatus of plants and bacteria combine atomically precise pigment-protein complexes with dynamic membrane architectures to control energy transfer on the 10-100 nm length scales. Recently, synthetic materials have integrated photosynthetic antennae proteins to enhance exciton transport, though the influence of artificial packing on the excited-state dynamics in these biohybrid materials remains unclear. Here, we use the adaptive Hierarchy of Pure States (adHOPS) to perform a formally exact simulation of excitation energy transfer within artificial aggregates of light harvesting complex 2 (LH2) with a range of packing densities. We find that LH2 aggregates support a remarkable exciton diffusion length ranging from 100 nm at a biological packing density to 300 nm at the densest packing previously suggested in an artificial aggregate. The unprecedented scale of these calculations also underscores the efficiency with which adHOPS simulates excited-state processes in realistic molecular materials. 
\end{abstract}

\twocolumn
\section{Main Text}
The ability to control energy transfer on the 10 - 100 nm length scale (i.e., the `mesoscale') is essential to designing new materials with applications in optoelectronics,  photocatalysis, and light harvesting.
The photosynthetic apparatus of plants and bacteria combines atomically precise structures of individual pigment protein complexes with a dynamic membrane architecture that has both inspired new light harvesting materials and stimulated advances in theoretical \cite{sumiTheoryRatesExcitationEnergy1999, scholesAdaptingForsterTheory2001, jangMultichromophoricForsterResonance2004a, yangInfluencePhononsExciton2002a, ishizakiUnifiedTreatmentQuantum2009a, rebentrostEnvironmentassistedQuantumTransport2009} and experimental \cite{brixnerTwodimensionalSpectroscopyElectronic2005, harelSingleShotGradientAssistedPhoton2011, dahlbergMappingUltrafastFlow2017a, arsenaultVibronicMixingEnables2020b, biswasCoherentTwoDimensionalBroadband2022} characterization of excitation energy transfer and charge separation. 
More recently, biohybrid materials that integrate photosynthetic proteins have performed remarkably well.\cite{reynolds_directed_2007, escalante_directed_2008, escalante_nanometer_2008, vasilev_reversible_2014, mancini_multi-step_2017,yoneda_photosynthetic_2020,kim_photosynthetic_2021}

Controlling excited-state processes in artificial assemblies of pigment-protein complexes requires understanding the mechanism of excitation transport on the mesoscale.  
Light harvesting complex 2 (LH2) from purple bacteria is a widely studied biological antenna protein \cite{mattioni21_041003, zheng_fully_2021, strumpfer_light_2009, scholes2000mechanism} that has been previously incorporated into artificial materials. \cite{escalante_directed_2008, reynolds_directed_2007}  
Early mechanistic studies of excitation energy transfer between LH2 proteins in biological assemblies suggest an incoherent mechanism of transport (i.e., excitations `hopping') between the dipole-allowed bright states. \cite{strumpfer_light_2009, jiminez_excitation_transfer_LH2} 
Recent simulations of two-dimensional LH2 aggregates in artificial assemblies have suggested the possibility of coherent transport on a relatively-long 500 fs timescale (using the single-D1 Davydov ansatz)\cite{zheng_fully_2021} or dark-state mediated transport arising between close-packed LH2 pairs (using a Lindblad master equation). \cite{mattioni21_041003} 
On the other hand, a generalized master equation simulation suggests a dominantly incoherent mechanism of transport even for very close-packed LH2 rings. \cite{jang_robust_2018} 
While generalized master equations, such as generalized F\"{o}rster theory, are accurate at biological inter-ring distances,\cite{strumpfer_light_2009} changes in aggregate packing can influence the appropriateness of different approximation schemes. 
Thus, assigning the mechanism of excitation transport in artificial LH2 aggregates, without imposing \textit{a priori} constraints, requires a formally exact method that can simulate excitation energy transport across a large number of bacteriochlorophyll.

While there exist many formally exact methods for calculating exciton dynamics, such as the multi-D1 Davydov ansatz,\cite{multi-D1} Hierarchical Equations of Motion (HEOM),\cite{tanimuraNumericallyExactApproach2020} Hierarchy of Pure States (HOPS),\cite{suessHierarchyStochasticPure2014} and quasi-adiabatic path integrals (QUAPI),\cite{makriImprovedFeynmanPropagators1991} they are limited to small aggregates by the rapid scaling of their computational expense with the number of simulated molecules. 
Recent advances in reduced scaling techniques - ranging from modular path integrals \cite{makriCommunicationModularPath2018, makriModularPathIntegral2018} to tensor-contracted methods \cite{borrelliExpandingRangeHierarchical2021, yanEfficientPropagationHierarchical2021, gaoNonMarkovianStochasticSchrodinger2022,somozaDissipationAssistedMatrixProduct2019, tamascelliEfficientSimulationFiniteTemperature2019, strathearnEfficientNonMarkovianQuantum2018b} and adaptive basis set techniques \cite{varveloFormallyExactSimulations2021} - raise the possibility of simulating excitation transport in mesoscale photosynthetic aggregates using formally exact methods, but such calculations have not been reported to date.

In this paper, we perform the first formally exact simulations of mesoscale excitation transport in LH2 aggregates using the adaptive Hierarchy of Pure States (adHOPS) method. \cite{varveloFormallyExactSimulations2021} Our calculations suggest LH2 aggregates can support a remarkable excitation diffusion length of up to 300 nm.

We model excitation energy transport using an open quantum system Hamiltonian \cite{schroter15_1}
\begin{equation}
    \label{eq:Ham_Full}
    \hat{H} = \hat{H}_{\textrm{S}} + \sum_{n,q} \kappa_{q_n} \hat{L}_n (\hat{a}^\dagger_{q_n} + \hat{a}_{q_n}) + \sum_{n,q} \hslash \omega_{q_n} (\hat{a}^\dagger_{q_n} \hat{a}_{q_n} + 1/2)
\end{equation}
where $\omega_{q_n}$ is the frequency of a harmonic oscillator with corresponding creation and annihilation operators $\hat{a}^\dagger_{q_n}$ and $\hat{a}_{q_n}$, and $\hat{L}_n$ is an operator ($\hat{L}_n = |n\rangle\langle n|$) that couples
the $n^{\textrm{th}}$ pigment to its independent environment described by bath modes $\{q_n\}$. The influence of the thermal environment on the electronic (i.e., system) energy levels is described by a correlation function
\begin{equation}
\label{eq:C_t}
C_n(t) = \frac{\hslash}{\pi}\int_0^\infty d\omega J_n(\omega) \big(\coth(\hslash \beta \omega /2) \cos(\omega t) - i \sin( \omega t)\big)
\end{equation}
which we decompose into a sum of exponentials called `correlation function modes' indexed by $j_n$
\begin{equation}
\label{eq:alpha_dl_coarse}
    C_n(t) = \sum_{j_n} g_{j_n} e^{-\gamma_{j_n} t/\hslash}. 
\end{equation}
where $\beta=\frac{1}{k_BT}$ is the inverse temperature and $J_n(\omega)$ is the spectral density.

We simulate the exciton dynamics using the Hierarchy of Pure States (HOPS), \cite{suessHierarchyStochasticPure2014} a formally exact solution to the open quantum system Hamiltonian (Eq. \eqref{eq:Ham_Full}). In HOPS, the full state of the system and bath is expressed as a collection of wave functions indexed by a vector $\vec{k}$, where $\vert \psi^{(\vec{0})}_t \rangle$ is the physical wave function and the remainder are referred to as `auxiliary wave functions.' The reduced density matrix for the system is given by an ensemble average ($\mathbb{E}[\cdot]$) over $N_{traj}$ wave function trajectories 
\begin{equation}
\rho_S = \mathbb{E}\left[\vert \psi^{(\vec{0})}(t; \mathbf{z}_t) \rangle \langle \psi^{(\vec{0})}(t; \mathbf{z}_t) \vert\right] 
\end{equation}
subject to a complex, stochastic noise $\mathbf{z}_t$ where components associated with individual thermal environments $z_{n,t}$ are defined by $\mathbb{E}[z_{n,t}]=0$,  $\mathbb{E}[z_{n,t} z_{n,s}]=0$, and $\mathbb{E}[z^*_{n,t} z_{n,s}]=C_n(t-s)$. For notational simplicity, we will refer to $|\psi(t,\mathbf{z}_t)\rangle$ as $|\psi_t\rangle$. The time-evolution of the HOPS wave functions is then given by
\begin{flalign}
\begin{aligned}
\label{eq:NormNonLinearHops}
\hslash \partial_t \vert \psi^{(\Vec{k})}_t \rangle 
=  \big(-i\hat{H}_S - \Vec{k} \cdot \Vec{\gamma} -\Gamma_t + \sum_{n} \hat{L}_{n} (z^*_{n,t}+ \xi_{n,t})\big)\vert \psi^{(\Vec{k})}_t \rangle \\ 
+ \sum_{n,j} k_{j_n} \gamma_{j_n} \hat{L}_{n}  \vert \psi^{(\Vec{k} -\Vec{e}_{j_n})}_t \rangle
\\- \sum_{n,j} (\frac{g_{j_n}}{\gamma_{j_n}})(\hat{L}^{\dagger}_{n} - \langle\hat{L}^{\dagger}_{n}\rangle_{t}) \vert \psi^{(\Vec{k}+\Vec{e}_{j_n})}_t\rangle
\end{aligned}
\end{flalign}
where 
\begin{equation}
\label{eq:memory_drift}
    \xi_{n,t} = \frac{1}{\hslash}\int_{0}^{t} ds C^{*}_{n}(t-s) \braket{L^{\dagger}_{n}}_s
\end{equation}
is a memory term that causes a drift in the effective noise, 
\begin{equation}
\label{eq:expectation_L}
    \langle\hat{L}^{\dagger}_{n}\rangle_{t} = \langle \psi^{(\vec{0})}_t \vert \hat{L}^{\dagger}_{n}\vert \psi^{(\vec{0})}_t \rangle,
\end{equation}
and
\begin{flalign}
\begin{aligned}
\label{eq:gamma_t}
    \Gamma_t = &\sum_{n} \braket{\hat{L}_{n}}_{t} \textrm{Re}[z^*_{n,t}+ \xi_{n,t}] \\
    - &\sum_{n,j}  \textrm{Re}\Big[\Big(\frac{g_{j_n}}{\gamma_{j_n}}\Big)\braket{\psi^{(\vec{0})}_t |\hat{L}^{\dagger}_{n}| \psi^{(\vec{e}_{j_n})}_t}\Big] \\
    + &\sum_{n,j}  \braket{\hat{L}^{\dagger}_n}_{t} \textrm{Re}\Big[\Big(\frac{g_{j_n}}{\gamma_{j_n}}\Big)\braket{\psi^{(\Vec{0})}_t | \psi^{(\Vec{e}_{j_n})}_t }\Big]
\end{aligned}
\end{flalign}
ensures normalization of the physical wave function. In these equations, $k_{j_n}$ is the ${j_n}^{\textrm{th}}$ element of the index vector $\vec{k}$, and $\Vec{\gamma}$ is a vector of the exponential factors of the correlation function modes. Here we employ the "triangular truncation" scheme, which restricts the hierarchy of auxiliary wave functions to a finite depth $k_{max}$, such that $\{\Vec{k} \in \mathbb{A}:  \sum_i k_i \leq k_{max}\}$, though other static filtering approaches have been proposed. \cite{zhangFlexibleSchemeTruncate2018}

The computational expense of HOPS scales poorly with system size because the number of auxiliary wave functions in the hierarchy increases rapidly with the total number of exponential modes defining the correlation functions. However, the physical wave function of a HOPS trajectory is localized in the basis of system states ($\mathbb{S}$) by the interaction with the thermal bath (`dynamic localization'), which induces a corresponding localization in the basis of auxiliary wave functions ($\mathbb{A}$).\cite{varveloFormallyExactSimulations2021} As a result, in extended aggregates, the vast majority of auxiliary wave functions are unoccupied at any single point in time. If unoccupied system states and auxiliary wave functions could be efficiently eliminated from the calculation, the computational expense of HOPS would not increase with system size (i.e., $\mathcal{O}(1)$ scaling) for sufficiently large aggregates.\cite{varveloFormallyExactSimulations2021}

The adaptive hierarchy of pure states (adHOPS) approach constructs a time-evolving reduced basis to provide a numerically tractable formulation of HOPS trajectories for large molecular aggregates.\cite{varveloFormallyExactSimulations2021} Every $u_t$ (`update time') femtoseconds, adHOPS constructs a reduced basis as a direct sum of reduced system and auxiliary bases ($\mathbb{S}_t\bigoplus\mathbb{A}_t$). 
The algorithm determines the error introduced by neglecting each basis element as an upper-bound on the Euclidean norm of the resulting displacement in the full derivative vector (Eq. \eqref{eq:NormNonLinearHops}). 
It then truncates basis elements in order of increasing error until user-defined error bounds for the hierarchy ($\delta_A$) and  then the state ($\delta_S$) bases are saturated. 
Previously, adHOPS has replicated the dynamics of HOPS in a linear chain with a dramatically reduced basis that demonstrated size-invariant (i.e., $\mathcal{O}(1)$) scaling for sufficiently large aggregates.\cite{varveloFormallyExactSimulations2021}

\begin{figure}[ht!]
	\includegraphics{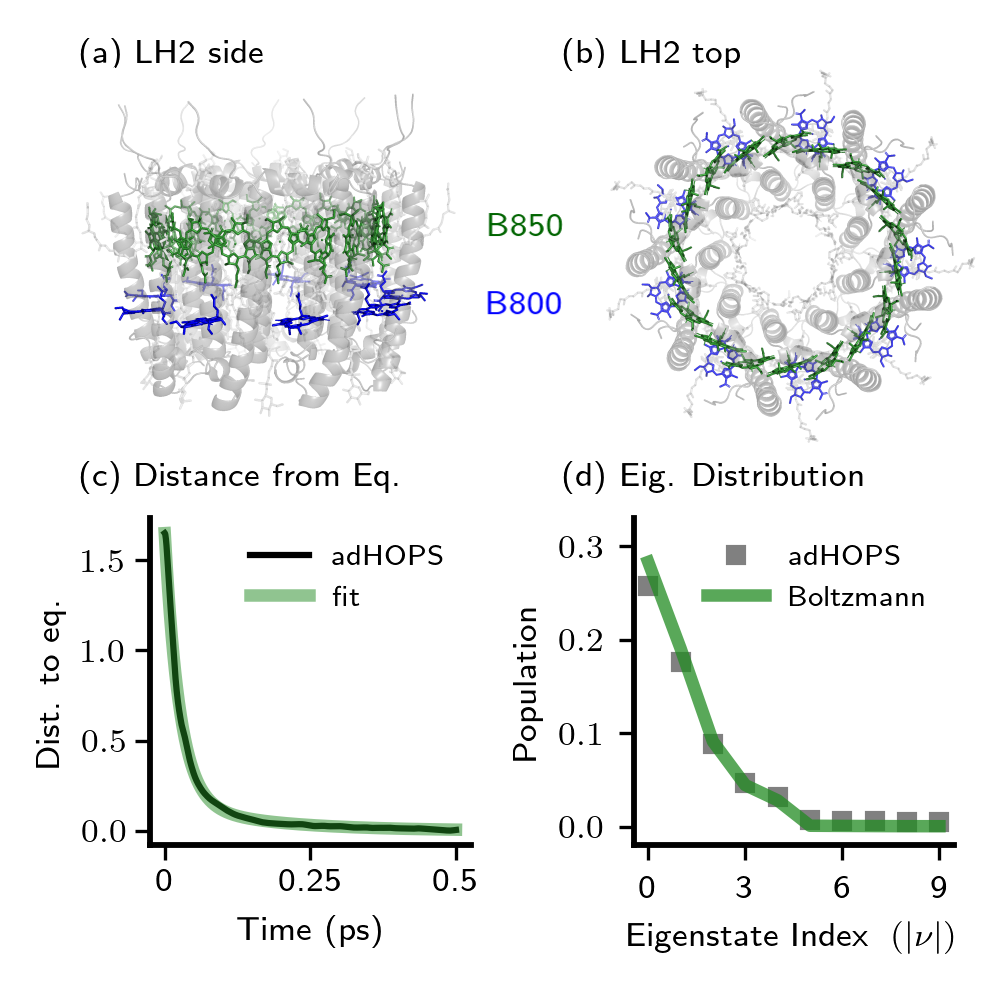}
	\caption{LH2 monomer. (a) Side view: the B850 ring (18 BChl) and the B800 ring (9 BChl) are shown in green and blue, respectively. (b) Top view. (c) The distance from equilibrium as a function of time (black) compared to a bi-exponential fit (green). (d) Boltzmann distribution (green line) compared to adHOPS equilibrium populations (grey squares) of  eigenstates ordered by absolute index ($\vert \nu \vert$).  $\sigma=160$ cm$^{-1}$ for site energy static disorder. Convergence parameters are given in Table S6 of the Supporting Information.}
	\label{fig:monomer_data}
\end{figure}

The version of adHOPS used for the calculations reported here is available in the MesoHOPS package version 1.1. \cite{mesohops_1_1} Additional details about both the algorithm and the calculation parameters are available in the Supporting Information.  We check the convergence of adHOPS results by running a series of adHOPS ensembles with increasingly strict parameters until the characteristic observables are within 2\% of the most exact calculations (see section S4 of the Supporting Information for details).

The LH2 monomer (Fig.~\ref{fig:monomer_data}a-b) contains two rings of bacteriochlorophyll (BChl) that absorb at 800 nm (B800 ring, blue) and 850 nm (B850 ring, green), respectively. The B800 ring is comprised of 9 widely-spaced and weakly-coupled BChls and funnels excitation to the B850 ring, which is responsible for transport between LH2 monomers. The tight packing of the 18 BChls (organized into $\alpha\beta$ pairs) in the B850 ring gives rise to strong electronic couplings and delocalized eigenstates.  Since inter-LH2 transport occurs predominately between B850 rings, \cite{hu_ritz} we neglect the B800 pigments found in the 1NKZ pdb structure. \cite{LH2_pdb}


We model the electronic energy levels of the B850 ring using previously established parameters.\cite{smythB800B850Coherence2015} In the electronic (`system') Hamiltonian, the vertical excitation energies (or `site energies') of the $\alpha$ and $\beta$ chlorophylls are 12690 cm$^{-1}$ and 12070 cm$^{-1}$, respectively, and the electronic couplings ($V_{n,m}$) within and between $\alpha\beta$ pairs are 307 cm$^{-1}$ and 237 cm$^{-1}$, respectively (see section S1 of the Supporting Information for full B850 Hamiltonian). In keeping with previous spectroscopic assignments, we also include static disorder on the site energy of each chlorophyll as Gaussian fluctuations with a standard deviation of $\sigma = 160$ cm$^{-1}$. \cite{smythB800B850Coherence2015, scholes2000mechanism} Finally, we model the thermal environment of each pigment with a Drude-Lorentz spectral density characterized by a reorganization timescale ($\gamma_{0_n}= 53$ cm$^{-1}$) and reorganization energy ($\lambda_n= 65$ cm$^{-1}$) \cite{smythB800B850Coherence2015} 
\begin{equation}
    \label{J_w}
    J_n(\omega) = 2\lambda_n\gamma_{0_n}\frac{\omega}{\omega^2 + \gamma_{0_n}^2}.
\end{equation}
The corresponding correlation function is composed of one high temperature mode, $k_{Mats}$ Matsubara modes, and an additional mode to ensure $\textrm{Im}[\alpha(0)]=0$ (see section S2 of the Supporting Information for details). We characterize the relaxation of the exciton population towards equilibrium using the 1-norm of a difference vector
\begin{equation}
    ||\underbar{P}(t)- \underbar{P}_{eq}||_1
\end{equation}
where $\underbar{P}(t)$ and $\underbar{P}_{eq}$ are the eigenstate population vector of the B850 ring at time $t$ and its equilibrium value, respectively. Using these parameters, the degenerate optically bright ($\nu=\pm 1$) states in the B850 ring relax to equilibrium on two timescales ($\tau_1 = 25 \,\,\mathrm{fs} \textrm{ and } \tau_2 = 150 \,\,\mathrm{fs}$, Fig.~\ref{fig:monomer_data}c), which agree with those found by global kinetic fits of B850 intraband relaxation in 2D electronic spectroscopy (50 fs and 150 fs). \cite{thyrhaugIntrabandDynamicsExciton2021} The equilibrium eigenstate populations (Fig.~\ref{fig:monomer_data}d) form a Boltzmann distribution, slightly perturbed by the electron-vibrational coupling.

\begin{figure*}
{\includegraphics{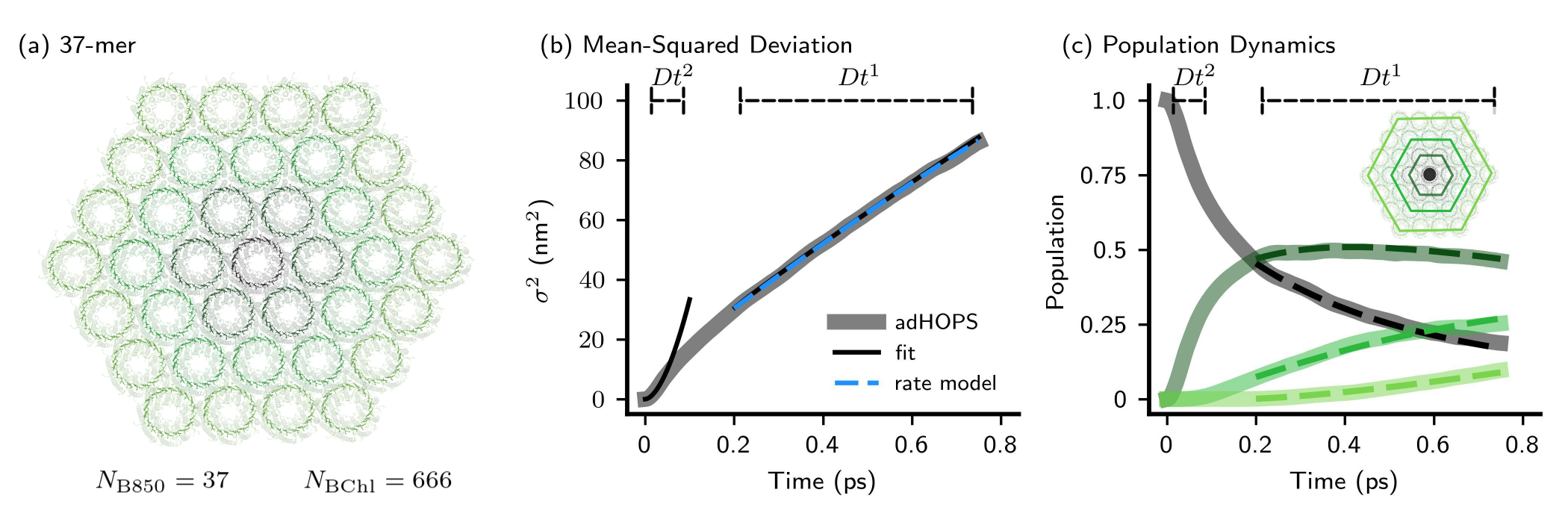}}
\caption{Exciton transport in a mesoscale aggregate of LH2. (a) Schematic of a hexagonally-packed B850 complex with three concentric shells. (b) Mean squared deviation (MSD) of the excitation from adHOPS dynamics (thick line), power-law fits (thin lines), and kinetic model (dashed line). (c) adHOPS (solid lines) and kinetic model (dashed lines) population dynamics for the donor and three concentric shells of B850 rings (shown in color-coded inset). $\sigma=160$ cm$^{-1}$ for site energy static disorder, and angular static disorder is given by randomly orienting all B850 rings in each trajectory. Convergence parameters are given in Table S6 of the Supporting Information.}
\label{fig:supercomplex_data}
\centering
\end{figure*}

We construct a hexagonally-packed supercomplex containing 37 B850 rings (666 BChl) organized as three concentric shells around a donor ring (Fig.~\ref{fig:supercomplex_data}a) with center-to-center ring distances ($R$) of 6.5 nm. This inter-ring distance represents the closest packing of LH2 proteins proposed for artificial aggregates. \cite{escalante_long-range_2010, mattioni21_041003} The inter-ring couplings $\Tilde{V}_{n,m}$ are calculated using the ideal dipole approximation 
\begin{equation}
    \Tilde{V}_{n,m} = \textrm{C} \frac{\vec{d}_n \cdot \vec{d}_m - 3(\vec{d}_n \cdot \vec{r}_{n,m})(\vec{d}_m \cdot \vec{r}_{n,m})}{R^3_{n,m}}
\end{equation}
where the coupling constant $\textrm{C}=348,000$ $\mathring{A}^3 \cdot cm^{-1}$, $\vec{d}_n$ is a unit vector along the direction of the transition dipole moment of pigment $n$, and $\vec{r}_{n,m}$ ($R_{n,m}$) is the unit vector (distance) between Mg atoms in pigments $n$ and $m$. In addition to the site energy static disorder included in each ring, we add an angular static disorder represented by randomly orienting the B850 rings in each trajectory. We find that static disorder in the site energies suppresses transport while angular disorder has negligible impact (see section S7 of the Supporting Information for details).

Exciton transport in LH2 aggregates exhibits two distinct regimes: a superdiffusive transport at short times and ongoing diffusive transport at times longer than 200 fs. To quantify transport in our adHOPS simulations, we calculate the mean-squared deviation (MSD, grey line, Fig.~\ref{fig:supercomplex_data}b) of the exciton distribution
\begin{equation}
    MSD(t) = \sum_n P_n(t)R_{0n}^2
\end{equation}
where $P_n$ is the population of the $n^{\textrm{th}}$ B850 ring and $n=0$ signifies the donor.
To characterize the mechanism of transport, we fit the MSD to a power law ($MSD(t) = Dt^\alpha$) in different time windows (black lines, Fig.~\ref{fig:supercomplex_data}b). From 0 to 50 fs, the MSD exhibits superdiffusive behavior ($\alpha = 1.74 \pm 0.11$). Between 50 and 200 fs, however, the MSD behavior shifts rapidly from a convex to a linear form, consistent with coherent transport giving way to diffusive transport on the timescale of vibrational reorganization ($\hslash/\gamma_{0_n} \approx 100$ fs). From 200 fs onward, the MSD remains diffusive ($\alpha = 0.95 \pm 0.04$) until the exciton reaches the boundaries of the complex. 

The 37-mer studied here is the minimal aggregate sufficient for simulating the turn over between coherent and diffusive transport without edge effects. 
By the onset of diffusive transport (200 fs), the exciton population summed over B850 rings forming the second concentric shell (inset, Fig.~\ref{fig:supercomplex_data}c) is nearly 10\% of the total population (medium green line, Fig.~\ref{fig:supercomplex_data}c), which would suppress further transport in a smaller aggregate.

\begin{figure}[ht!]
	\includegraphics{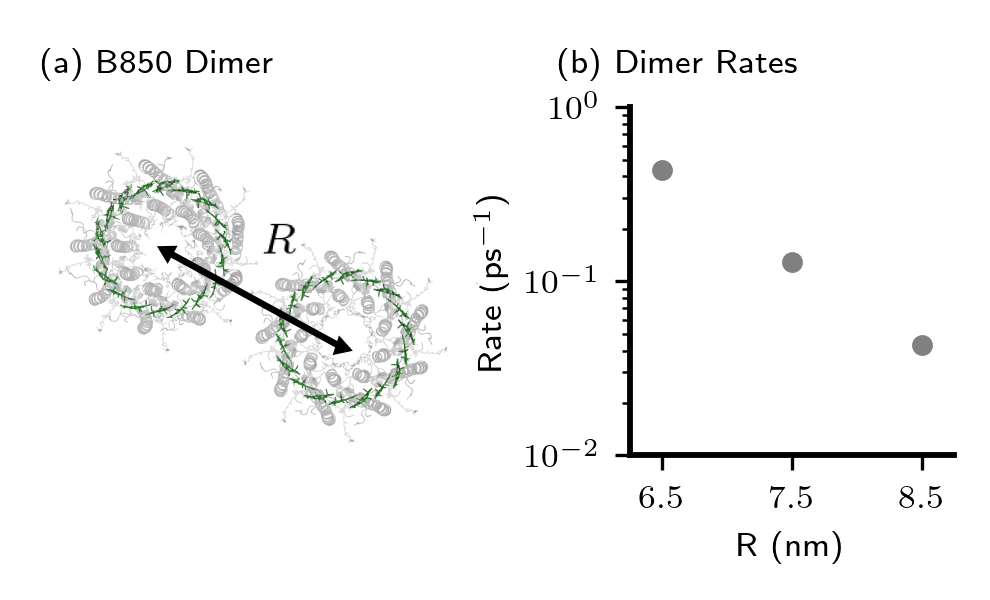}
	\caption{LH2 dimer rates. (a) Schematic of a B850 dimer with a center-to-center separation distance of $R$. (b) Transport rates for an LH2 dimer with an inter-ring separation of $R$. In all cases, $\sigma=160$ cm$^{-1}$ for site energy static disorder, and angular static disorder is given by randomly orienting both B850 rings in each trajectory. Convergence parameters are given in Table S6 of the Supporting Information.}
	\label{fig:bright_vs dark}
\end{figure}

The later-time diffusive transport of the 37-mer is reproduced by a kinetic model characterized by a single rate of transport. We fit the population dynamics of a B850 dimer (Fig.~\ref{fig:bright_vs dark}a) with center-to-center distance $R=6.5$ nm to a single rate, $\kappa_{dimer} = 0.44 \pm 0.04$ ps$^{-1}$ (see section S6 of the Supporting Information for details). 
We time-evolve the vector of B850 ring populations ($\underbar{P}(t)$) according to
\begin{equation}
\label{eq:rate_model}
    \dot{\underbar{P}}(t) = \underline{\underline{K}} \; \underbar{P}(t),
\end{equation}
where the rate matrix ($\underline{\underline{K}}$) connects nearest-neighbor rings via symmetric transport rates equal to $\kappa_{dimer}$. 
Starting from 200 fs, the population dynamics of the kinetic model (dashed lines) reproduce the results of adHOPS calculations (solid lines, Fig.~\ref{fig:supercomplex_data}c). 
This ensures that the kinetic model also reproduces the MSD dynamics of diffusive exciton transport in the LH2 aggregate (dashed line, Fig.~\ref{fig:supercomplex_data}b). We have also explored the possibility of the dark-state shelving mechanism recently proposed for close-packed LH2,\cite{mattioni21_041003}  but found that inter-ring transport is dominated by coupling between bright states (see section S8 of the Supporting Information for details). 

\begin{table*}
\caption{LH2 exciton diffusion lengths in a hexagonally-packed lattice}
\label{tab:diffusion_param}

\begin{tabular}{ |c|c|c|c|c| } 
 \hline
 $R$ (nm) & method & D (nm$^2$/ps$^\alpha$)& $\alpha$ & \hphantom{sp} $L_d$ (nm)\hphantom{sp} \\
 \hline
 \multirow{2}{*}{6.5}  & \hphantom{sp}rate model\hphantom{sp} & $110 \pm  10.$  & 1 & $330 \pm 15$ \\ 
 \cline{2-5}
    & adHOPS & $100 \pm  10.$  & $0.95 \pm 0.044$ & $270 \pm 54$ \\ 
 \hline
 7.5 & rate model & $43 \pm  2.5$ & 1 & $210 \pm 6$ \\
 \hline
 8.5  & rate model &  $19 \pm  1.0$ & 1 & $140 \pm 4$ \\

 \hline
\end{tabular}

 \vspace{0.1in} The uncertainty represents the 95\% confidence interval.
\end{table*}


 






The excitation diffusion length ($L_{d}$) in LH2 aggregates predicted by our kinetic model greatly exceeds that of prototypical organic semiconductors, even for the relatively loose packing associated with biological membranes. In Fig.~\ref{fig:bright_vs dark}b, we show the dimer rates of transport as a function of the packing distance ($R$) from 6.5 nm suggested in some artificial materials to 8.5 nm found in biological membranes. 
Table \textbf{\ref{tab:diffusion_param}} reports the excitation diffusion length in the kinetic model of an infinite aggregate (see section S9 of the Supporting Information) determined by 
\begin{equation}
    \label{diffusion_length}
    L_d^2 = 6R^2\kappa_{dimer}\tau \equiv D\tau,
\end{equation}
where $R$ is the packing distance, $\kappa_{dimer}$ is the LH2 dimer rate, and $\tau$ is the lifetime of the exciton (assumed to be 1 ns).\cite{escalante_long-range_2010}
At a biologically relevant inter-ring distance of $R=8.5$ nm, our excitation diffusion length ($L_d = 140$ nm) is consistent with a previous order of magnitude estimate \cite{jang2015molecular} and greatly exceeds that expected for a prototypical organic semiconductor ($<30$ nm). \cite{sajjad2020_OSC, firdaus2020_OSC} 
Moreover, our rate of transport at $R=8.5$ nm is smaller than some previous estimates (e.g., 0.241 ps$^{-1}$ in Ref. \onlinecite{jang_robust_2018}), suggesting other LH2 Hamiltonians proposed in the literature could support even longer excitation diffusion lengths.

We conclude that LH2 aggregates support exciton transport on length scales  exceeding those of similar artificial and natural materials. 
Excitation transport in LH2 aggregates exhibits a brief ballistic period ($<50 \textrm{ fs}$) followed by diffusive transport mediated by bright-state coupling that supports long-range exciton diffusion even at biological packing distances ($R=8.5$ nm, $L_d = 140$ nm). 
We note that LH2 complexes, despite belonging to the evolutionarily `early' anoxygenic purple bacteria, support a surprisingly large $L_d$ compared to that previously calculated for the oxygenic photosystem II membrane ($L_d = 50$ nm),\cite{amarnath_multiscale_2016, bennett_energy-dependent_2018} suggesting that the need for regulation and not long-range excitation energy transport has driven the design of antenna complexes in higher plants. 
Moreover, in artificially close-packed LH2 aggregates, our calculations suggest the exciton diffusion length can reach up to $300$ nm, a full order of magnitude larger than a prototypical organic semiconductor, making LH2 a promising antenna system for biohybrid materials.

In this paper, we have reported a formally exact simulation of exciton dynamics in a 
mesoscale photosynthetic aggregate consisting of an unprecedented 37 LH2s with a total of 666 bacteriochlorophylls. 
While some previous studies have simulated large LH2 aggregates using approximate methods, the current calculations provide an important benchmark result where the accuracy of the simulation is limited only by the parameterization of the model Hamiltonian. 
Our results also demonstrate that adHOPS is suitable for calculations of realistic mesoscale molecular materials to address mechanistic questions about excited-state processes.
We expect the continued development of adHOPS will enable a new generation of simulations capable of probing larger and more complex materials to reveal new strategies for controlling excited-state processes.

\begin{acknowledgement}
The authors thank Bailey Raber and Mohamed El Refaiy for assistance working with the protein structure and the electronic Hamiltonian for the extended aggregates, Julian 
Schmidt for helpful discussions and testing a preliminary version of the adHOPS code, as well as Tarun Gera for editing and review. LV, BC, and DIGB acknowledge support from the Robert A. Welch Foundation (Grant N-2026-20200401). JKL acknowledges support from a Moody Fellowship. 
OK acknowledges funding by the Deutsche Forschungsgemeinschaft  - SFB 1477 "Light-Matter Interactions at Interfaces", project number 441234705”.
\end{acknowledgement}

\begin{suppinfo}
Methodological details, all calculation parameters, convergence studies, analysis of dark/bright state mechanisms, and comparison of different types of static disorder (PDF)
\end{suppinfo}

\bibliography{achemso-demo}

\end{document}